\documentclass{emulateapj}
\usepackage{epsf}
\usepackage{apjfonts}
\usepackage{xspace}
\usepackage{amsmath}
\usepackage{framed}
\usepackage{units}

\newcommand{\angstrom}{\text{\normalfont\AA}}

\def\dim#1{\mbox{\,#1}}

\def\hide#1{}

\begin{document}

\title{Cosmic Reionization On Computers. Ultraviolet Continuum Slopes and Dust Opacities in High Redshift Galaxies}

\author{Zimu Khakhaleva-Li\altaffilmark{1} and Nickolay Y.\ Gnedin\altaffilmark{2,3,4}}
\altaffiltext{1}{Department of Physics, The
  University of Chicago, Chicago, IL 60637 USA; zimu@uchicago.edu}
\altaffiltext{2}{Particle Astrophysics Center, Fermi National Accelerator Laboratory, Batavia, IL 60510, USA; gnedin@fnal.gov}
\altaffiltext{3}{Kavli Institute for Cosmological Physics, The University of Chicago, Chicago, IL 60637 USA;}
\altaffiltext{4}{Department of Astronomy \& Astrophysics, The
  University of Chicago, Chicago, IL 60637 USA} 

\begin{abstract}
We compare the properties of stellar populations of model galaxies from the Cosmic Reionization On Computers (CROC) project with the exiting UV and IR data. Since CROC simulations do not follow cosmic dust directly, we adopt two variants of the dust-follows-metals ansatz to populate model galaxies with dust. Using the dust radiative transfer code Hyperion, we compute synthetic stellar spectra, UV continuum slopes, and IR fluxes for simulated galaxies. We find that the simulation results generally match observational measurements, but, perhaps, not in full detail. The differences seem to indicate that our adopted dust-follows-metals ansatzes are not fully sufficient. While the discrepancies with the exiting data are marginal, the future JWST data will be of much higher precision, rendering highly significant any tentative difference between theory and observations. It is, therefore, likely, that in order to fully utilize the precision of JWST observations, fully dynamical modeling of dust formation, evolution, and destruction may be required.
\end{abstract}


\section{Introduction}
\label{sec:intro}

Many astronomers would agree that the launch of the Hubble Space Telescope (HST) opened a golden age in astronomy; for studies of reionization that gold is encrusted with diamonds. Various HST deep fields contain over 1500 galaxies beyond $z\sim6$ and reaching all the way back to $z\sim10$ \citep{rei:biot15}, directly probing what is likely to be dominant reionization sources \citep[but see][for an alternative viewpoint]{rei:mh15}. The future of the field is even brighter, as not only James Webb Space Telescope (JWST) will greatly expand samples of high redshift galaxies and push the observational limits to significantly fainter levels, but several other observational ventures, from ALMA observing dust re-emission from these galaxies to radio observations of redshifted $21\dim{cm}$ line and optical and infrared (OIR) spectra from $z\sim6$ quasars with 30-meter class telescopes will complement JWST with extensive constraints on the effect of high redshift galaxies on their environments and general IGM.

The prospects of observational bonanza force theorists to revise their tools - there is little doubt that future data will blow all existing theoretical models out of the water (or, rather, out of the list of valid theories). Numerical simulations of reionization, in particular, are currently going through a genuine revival, breaking all records in computational complexity, spatial and mass scales modeled, and physical fidelity \citep[c.f.][for a grossly incomplete list]{rei:ima14,newrei:snr14,newrei:nrs15,newrei:hdp14,newrei:dwm15,newrei:tcm15,newrei:bsv15}. Our own contribution to that effort is the ``Cosmic Reionization on Computers'' (CROC) project \citep{ng:g14}, which we describe below.

While numerical challenges for modeling reionization are formidable, they already fall within the capabilities of largest existing supercomputers. Hence, a particular emphasis in the future simulation work will be in improving fidelity of physical models for star formation and feedback (and, possibly, quasar formation). Such models will necessarily be phenomenological, and, hence, need to be calibrated against the observational data. A combination of JWST and ALMA observations of high redshift galaxies will be a particular strong test on any star formation and feedback recipes used in future simulations - they will not only constraint star formation histories of high redshift galaxies, but will also measure (or limit) re-emission of stellar radiation by dust in the sub-millimeter spectral window. Hence, theorists' task is to bring both on-the-fly modeling of star formation and feedback and post-processing of simulations to the level, compatible in detail and precision to the future observations.

As a first step in implementing this program for the CROC project, we describe in this paper our work on modeling radiative transfer of ultraviolet (UV) radiation from young stars through the ISM of simulated galaxies. The goal of this effort is two-fold: first, to make accurate predictions for the luminosities and spectral shapes of galaxy emission in the observed OIR range for predicting JWST observations, and second, to compute spectra of radiation re-emitted by dust for making predictions for ALMA.

There exist several codes capable of modeling radiative transfer through dusty media such as Sunrise \citep{zimu:sunrise1,zimu:sunrise2}, SKIRT \citep{zimu:SKIRT}, and RADMC-3D \citep{zimu:RADMC}. For our goal we find most convenient to use the code \textsc{Hyperion} \citep{zimu:Hyperion}, because its internal data structures are best-matched to the data structures used by our numerical simulation code, Adaptive Refinement Tree \citep[ART, ][]{misc:k99,misc:kkh02,sims:rzk08,ng:g14}, and because \textsc{Hyperion} includes all the functionally that is needed for our purposes (absorption and scattering of UV photons by dust, re-emission of sub-millimeter photons by dust). 

In the following, we briefly describe CROC simulations in \S 2, discuss radiative processes and our use of \textsc{Hyperion} in \S 3, and present our results in \S 4.

\section{Simulations}
\label{sec:sims}

Theoretical models for high redshift galaxies come from a set of three CROC simulations in $40h^{-1}\dim{Mpc}$ boxes (simulations B40.sf1.uv2.bw10.A-C from \citet{ng:g14}), with the exception of $z\approx5$ epoch, where $40h^{-1}\dim{Mpc}$ simulation data are unavailable. At $z\approx5$ we use six CROC simulations in $20h^{-1}\dim{Mpc}$ boxes (simulations B20.sf1.uv2.bw10.A-F from \citet{ng:g14}), which are equivalent in volume to 0.75 of a single $40h^{-1}\dim{Mpc}$ simulation. With spatial resolution reaching down to $125\dim{pc}$ in proper units at $z=6$, these simulations resolve radial structure of galactic disks, but not their thicknesses. This resolution is well matched to the scales on which the empirical Kennicutt-Schmidt relation -- used as a star formation recipe in the CROC simulations -- is well established. These are also the scales where the adopted stellar feedback model (``blastwave'' or ``delayed cooling'' model) has been shown to work well \citep{sims:sdqg09,sims:gbmb10,sims:atm11,sims:bsgk12,sims:aklg13,sims:sbmw13}. With these two main physical ingredients, CROC simulations reproduce the whole time evolution of the galaxy luminosity function up to $z\approx6$, i.e.\ the model galaxies in the simulations live in correct dark matter halos and have correct star formation histories. 

The three independent realizations of the $40h^{-1}\dim{Mpc}$ box and the six independent realizations of the $20h^{-1}\dim{Mpc}$ box used in this paper rely on the ``DC mode formalism'' \citep{sims:p97,sims:s05,ng:gkr11} to properly account for the fluctuation power outside the simulation box size. As the result, they provide an unbiased sample of galaxies up to the scales comparable to the simulation box size, and differences between the realizations can be used to estimate cosmic variance.

We analyze snapshots at $z\approx5$ and include the results here for completeness and because of the relative abundance of observational data at this redshift. However these results are not to be taken with a similar degree of fidelity as those at higher redshifts. The spatial resolution of all simulations that are used in this paper is fixed in comoving units, which implies that the proper spatial resolution degrades as the simulation evolves to lower redshifts. This degradation of spatial resolution results in a mismatch between the simulated galaxy UV luminosity function at $z=5$ and the observational data for galaxies fainter than about $M_{\rm AB}=-20$.

\section{Radiative Processes}
\label{sec:rad}
\subsection{Stellar Spectra}
Our simulations adopt an empirical sub-grid model of star formation \citep{ng:g14}, that adopts an approximately linear correlation between the star formation rate surface density $\Sigma_{\mathrm{SFR}}$ and the molecular gas surface density $\Sigma_{\mathrm{mol}}$ \citep{ng:g14}\begin{equation}
\Sigma_{\mathrm{SFR}}=\frac{\Sigma_{\mathrm{mol}}}{\tau_{\mathrm{SF}}},
\end{equation} where a fiducial value of $\tau_{\mathrm{SF}}=1.5\mathrm{Gyr}$ is assumed for the molecular gas depletion time scale. The delayed cooling model \citep{sims:sdqg09,sims:gbmb10,sims:atm11,sims:bsgk12,sims:aklg13,sims:sbmw13} is implemented as the stellar feedback recipe in the simulations to reduce star formation rate and therefore produce galaxies with relatively realistic properties for most of the redshift range of interest.

The stars are formed with a fixed Kroupa IMF, and the ionizing part of the spectrum comes from Starburst99. The entire spectrum of each stellar particle from ionizing UV to near infrared (from $900 \angstrom$ to $9990 \angstrom$ with $10 \angstrom$ resolution) which contains most of a star's power in electromagnetic radiation is computed using the Flexible Spectral Population Synthesis (FSPS) code \citep{misc:cgw10,misc:cg10}. This enables us to accurately estimate the bolometric luminosity and to compute dust re-emission fluxes from our dust model described below that can be directly compared to observations.

\subsection{Dust Model}
\label{subsec:dust}
The dust-to-gas ratio in the ISM is a result of a complex dynamical equilibrium between dust production through metal accretion onto seed grains in the stellar winds of asymptotic giant branch (AGB) stars and ejected by type II supernovae (SNII), and dust grains destruction through UV photodesorption and photolysis as well as collisions and bombardments in hot gas \citep{ism:i03,ism:d09,ism:i11,zimu:Asano2013,zimu:Feldmann2015}.

As most of the dust in the ISM is possibly not stardust, but is instead produced through metal accretion \citep{ism:d09}, the canonical assumption is for the dust-to-gas ratio to scale linearly with the gas metallicity \citep{zimu:leroy11,zimu:Feldmann2015}. Due to the relatively low galaxy metallicities at the redshifts of interest for reionization, the SMC dust model is generally assumed to represent better the dust properties in these galaxies compared to the Milky Way dust model. In fact, most of the galaxies at $z\sim 5 - 6$ in our simulations have metallicities close to that of the SMC \citep{ng:g14}. For this reason and also for consistency with previous CROC analyses, we adopt an isotropic SMC dust model \citep{misc:wd01} with density before sublimation given by the canonical assumption with a fiducial dust-to-metal ratio of 50\% \citep{ism:i11}. In any case, tests have revealed that adopting SMC, LMC, or MW dust model generally does not yield significantly different results \citep{ism:ddbg07}. The ISM of high redshift galaxies have been observed to possess dust-to-metal ratios close to that of the Local Group \citep{zimu:Zafar2013,zimu:Sparre2014}, providing support to the adoption of our fiducial dust-to-metal ratio.

The canonical assumption above is reasonable, if dust accretion in the ISM is the main mechanism of dust production. However, it does not account for dust destruction and is therefore physical only in the absence of processes destroying dust grains or if such processes operate on a similar timescale as dust accretion \citep{zimu:Feldmann2015}, in which case dust destruction simply lowers the dust-to-metal ratio. It is believed that the physical processes most efficient in destroying cosmic dust are supernova shocks, where dust grains are more likely to collide due to the high density and are also sputtered due to the high temperature \citep{ism:i11,ism:i03,zimu:Asano2013,zimu:Feldmann2015}. Just as in the case of dust ejection, the effect of SNII on dust destruction is dominant compared to that of SNIa \citep{zimu:Asano2013}. The former are associated with starburst and are therefore primarily found in HII regions where the ISM has been ionized by O and B stars produced by the active star formation. Thus we can use HII regions in the simulations as tracers of type II supernovae shocks. Moreover, the conditions in HII regions themselves are conducive to dust destruction through sputtering by hot gas and UV photolysis by O and B stars. Thus, to account for dust destruction physically without dynamically monitoring supernova shocks in the simulations, we introduce an instant sublimation model in which all dust grains in the ionized gas are instantly sublimated. Given the relatively short lifetime of dust grains, $\tau_{\mathrm{D}} \sim 100 \mathrm{Myr}$ \citep{ism:i11} on average, and the fact that the lifetime has to be even shorter significantly in HII regions compared to the average, the instant sublimation model represents a good compromise between computational complexity and physical accuracy.

The above models of dust production and destruction are based on current observational constraints and/or physically reasonable simplifications, and have been adopted to represent quasi-static equilibrium states. They allow us to analyze the effects of dust in our simulations without dynamically incorporating dust evolution models into the them. We plan to introduce dynamical models of dust formation into our simulations in future works.

\subsection{Radiative Transfer}
\label{subsec:radtrans}
\begin{figure}[htbp]
\includegraphics[width=1\hsize]{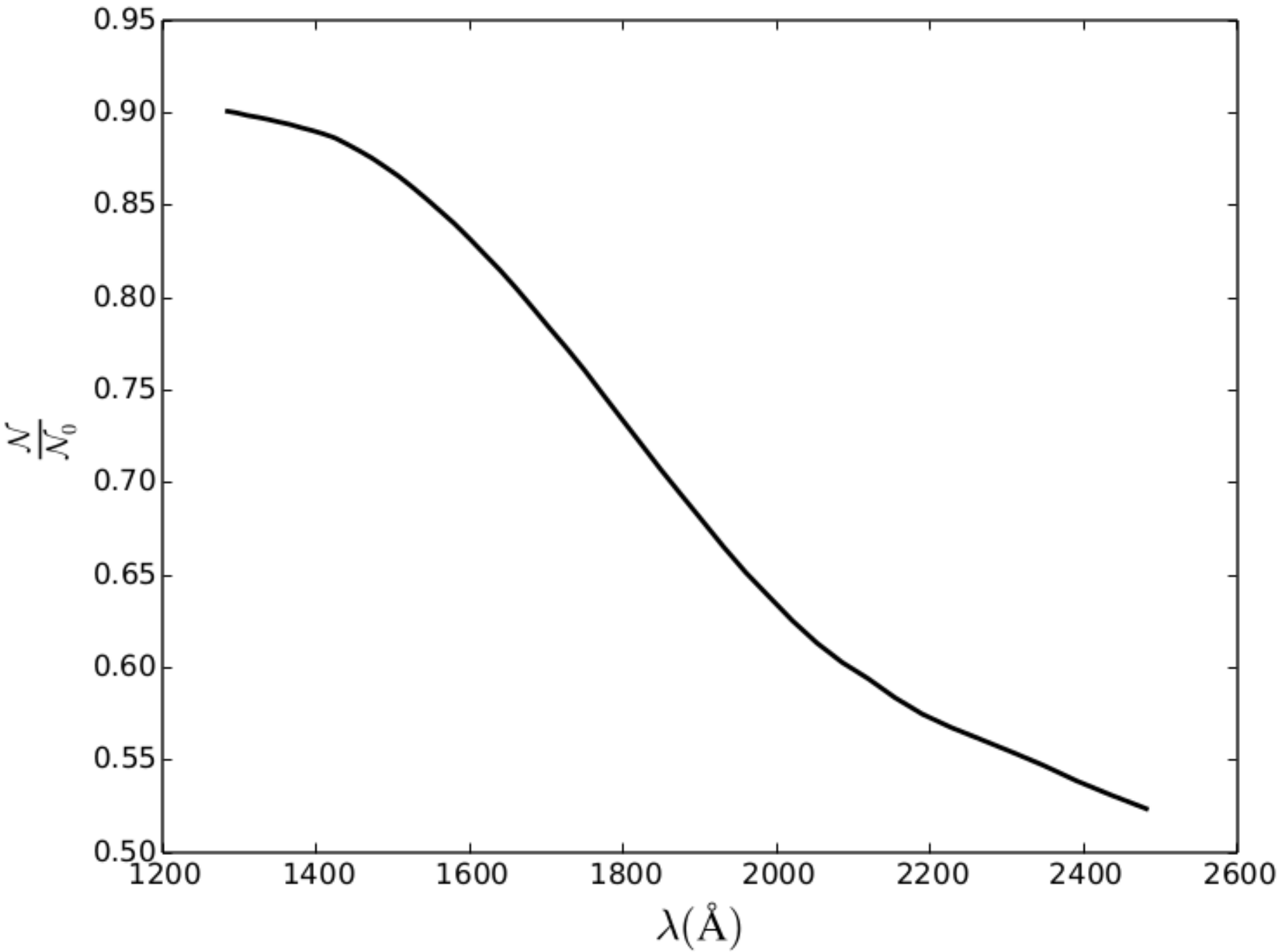}
\caption{Ratio of the effective dust column density with SMC dust albedo over that computed without elastic scattering as a function of wavelength for a single stellar particle. The latter is a completely flat line as expected. Instant dust sublimation model is used.}
\label{fig:N}
\end{figure}
The total dust optical depth of a galaxy is the summation over individual stars, \begin{equation}
\label{eq:taustar}
\tau(\lambda)=\sigma_{\mathrm{D}}(\lambda)\sum\limits_i \mathcal{N}_i,
\end{equation} where $\sigma_{\mathrm{D}}(\lambda)$ is the dust UV absorption cross-section and encompasses all the wavelength dependence of the dust optical depth, and $\mathcal{N}_i$ is the dust column density of the $i$th star. The first CROC papers \citep{ng:g14} adopt a simple dust obscuration model which includes only direct line-of-sight dust extinction, with the summation over stars in eq. \eqref{eq:taustar} replaced with an effective dust column density, 
\begin{equation}
\label{eq:taugal}
\tau(\lambda)\approx\sigma_{\mathrm{D}}(\lambda)\mathcal{N}_{\mathrm{eff}}.
\end{equation}

The assumption that $\mathcal{N}_{\mathrm{eff}}$ is wavelength independent, which is a direct consequence of the line-of-sight treatment, is physical only when elastic scattering of UV photons on dust particles is negligible, or when the dust albedo has little wavelength dependence. Neither of these is true for the SMC dust model that we use here, which has wavelength dependent albedo ranging from $0$ to close to $0.8$ \citep{misc:wd01}. Thus, a simple analytical dust obscuration model is insufficient, and a more sophisticated numerical radiative transfer treatment, taking into account both elastic and inelastic scatterings of UV photons from dust particles, is necessary to realistically model the effects of dust on star light in the simulations.

The numerical tool we have chosen to utilize is the Monte-Carlo dust-continuum radiative transfer code \textsc{Hyperion} \citep{zimu:Hyperion}. With native support for the octree grids utilized in our simulation sets, it enables radiative transfer post-processing on snapshots from the simulation sets with optimal resolution. 

Using this tool, we are able to explicitly verify as a preliminary test that the simple analytical treatment is inadequate. For simplicity and to eliminate potentially interfering factors, in this test we select a single actual stellar particle from one of our simulations. We also approximate the SMC dust optical depth as a power law \citep{misc:wd01,ng:gkc08,ng:g14}, 
\begin{equation}
\sigma_{\mathrm{D}}(\lambda)=\sigma_{\mathrm{D},1500}\left(\frac{1500\mathrm{\angstrom}}{\lambda}\right)^{1.1}.
\end{equation}
The instant dust sublimation model, as described in subsection \ref{subsec:dust}, is implemented. With only one stellar particle, eq. \eqref{eq:taugal} is exact with $\mathcal{N}_{\mathrm{eff}}\rightarrow\mathcal{N}$. We execute the radiative transfer to obtain the flux of this stellar particle with dust obscuration, which gives us the optical depth $\tau(\lambda)$ and therefore $\mathcal{N}$ from eq. \eqref{eq:taugal}, providing a direct test of the wavelength dependence of $\mathcal{N}$. We have observed that when the albedo $\alpha\equiv0$, $\mathcal{N}_0$ is indeed independent of wavelength as expected. However, figure \ref{fig:N} shows that $\mathcal{N}$ rises at wavelengths where the albedo is lower once the SMC albedo \citep{misc:wd01} is accounted for, indicating more absorption by inelastic scattering due to reduced probability of elastic scattering. The result is a clear indication of the competition between the two physical processes - inelastic scattering parametrized by the dust optical depth, and elastic scattering parametrized by the dust albedo. When an interaction between an UV photon and dust particle takes place, the two parameters determine the relative probability of the process to take place. At wavelengths where the albedo is higher, the light path is likely less direct and longer due to multiple elastic scatterings. But the decreased probability of an absorption at each interaction more than compensates for this effect and reduces extinction overall. This fact is unambiguously demonstrated by the fact that $\mathcal{N}$ is everywhere lower when elastic scatterings are present compared to the $\alpha\equiv0$ case. It is conceivable that part of this reduction could be due to effects of the potentially complex geometrical distribution of metals (and therefore dust) in the galaxy. However, the same effect is also observed when dust sublimation is turned off, thereby drastically altering the geometrical distribution of dust. This indicates that the competition between inelastic and elastic scatterings is principally responsible for this overall difference in $\mathcal{N}$. We have shown that the effects of utilizing the SMC dust are qualitatively different from those given by a reflection-less but otherwise identical dust as assumed in the simple dust obscuration model, and a numerical treatment of radiative transfer through dusty regions in the galaxies of our simulation sets is indeed necessary to ensure physical results.

\section{Results}
\label{sec:res}
\subsection{Galaxy Selection}
Due to practical limitations on computational resources, a random sample of 30 galaxies are selected in each snapshot for each dex in stellar mass, except when the mass bin in the snapshot contains fewer than 30 galaxies, in which case all galaxies are selected and the resultant difference in sample weight taken into account in later analyses.

\textsc{Hyperion} requires a strict octree grid structure. Since CROC simulations are performed with the oct-bases AMR code, it is natural to use the actual simulation grid as the radiative transfer grid for \textsc{Hyperion} - such setup preserves the exact spatial resolution of the simulation. However, some of simulated galaxies do not entirely fit into a single simulation root cell (either because they are too large or because they are located too close to the cell edge). Hence, we combine 8 adjacent cells of the simulation root grid into a ``supra-root'' cell, with 2 root cells or $312.5h^{-1}$ comoving kpc on a side, so that the center of the supra-root cell is the closest possible to the center of the galaxy. Such ``supre-root'' cells are sufficiently large to encompass all galaxies in the snapshots of our simulations sets even at the lowest redshift of interest $z\sim5$. 

With this approach, we also include in the supra-root cell all the satellites of a central galaxy. Hence, hereafter we only consider isolated galaxies, and combine satellites with their central galaxies for the purpose of measuring luminosity functions. Radiative transfer post-processing with \textsc{Hyperion} is thus performed individually for each galaxy using its ``supra-root'' cell as the root of the octree.

\subsection{Ultraviolet Attenuation}

One way of characterizing the effects of dust in galaxies is by observing extinction of starlight due to dust grains absorbing photons with wavelengths similar to their sizes or smaller. In particular, dust destruction by photodesorption and photolysis due to UV photons absorbed by grains plays a significant role in the dynamics of dust production and destruction in the ISM. Thus, for high-redshift galaxies, observing UV attenuation provides an important means of obtaining information about dust grains in the ISM. This information is accessed from our simulation snapshots by utilizing the capability of \textsc{Hyperion} to perform monochromatic radiative transfers \citep{zimu:Hyperion}. Starlight attenuation due to dust grains is computed at 50 wavelengths distributed uniformly from $1300\angstrom$ to $1700\angstrom$, consistent with previous work from CROC \citep{ng:g14}. To compare with existing observational data, we compute the AB magnitude UV luminosity functions, UV continuum slopes, and UV optical depths for SMC dust both with and without the instant sublimation model implemented, as well as the original starlight in the absence of dust (when applicable). The first and last quantities are computed at $1500\angstrom$. 

\begin{figure*}[ht]
\centering
\includegraphics[width=0.85\textwidth]{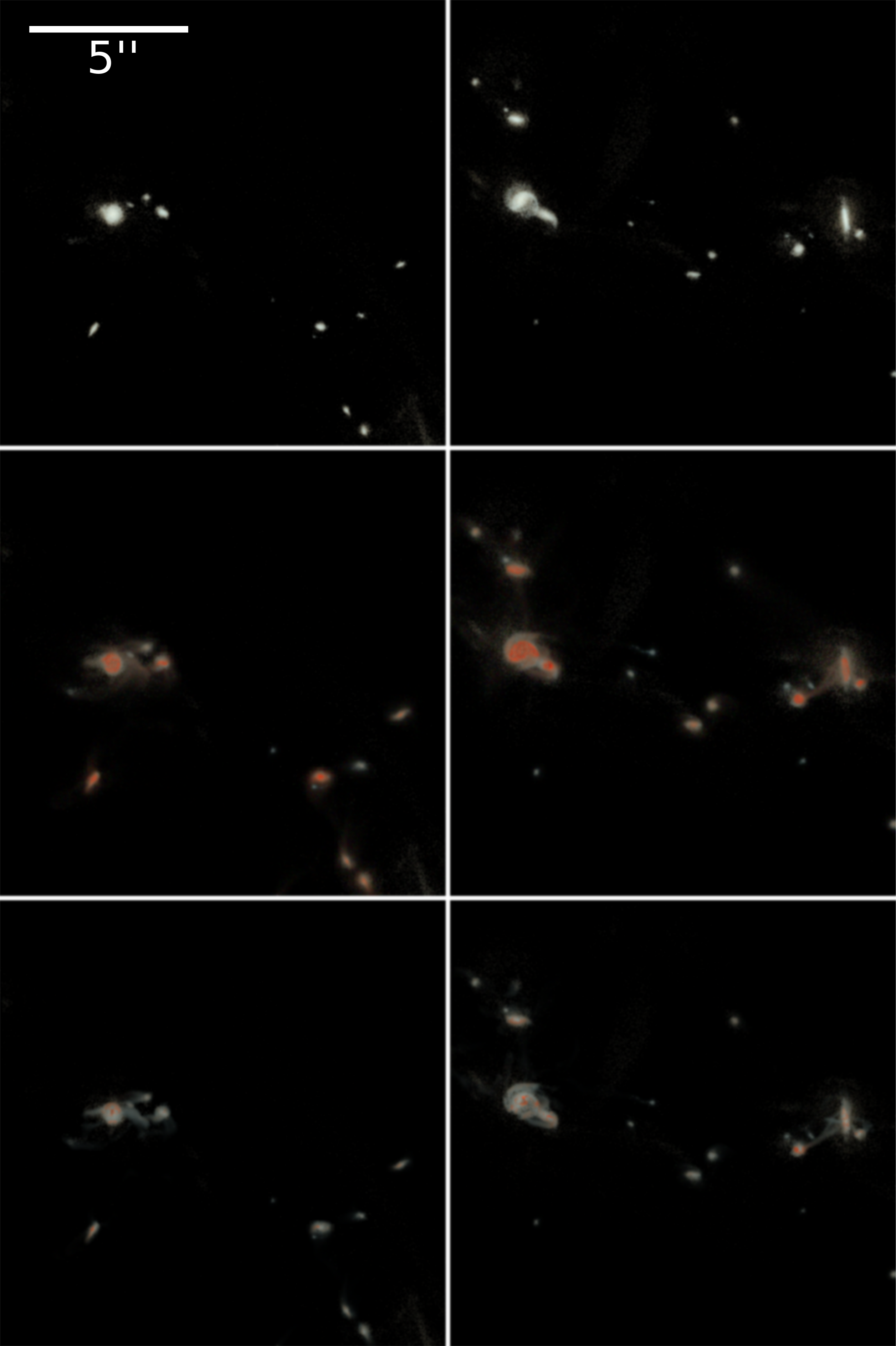}
\caption{Ultraviolet images of two most massive galaxies in realization C of our simulation sets at $z\sim6$, produced with the scaling from \citep{zimu:RGB} with $F(x)=\sqrt{\sinh^{-1}{(x/x_c)}}$ stretch, where $x_c=\bar{x}/8$ is an arbitrary choice that appears to maximally enhance galactic structures, and Gaussian smoothed to twice the pixel scale of the short-wavelength FPAs of NIRCam on JWST, at 0.064" \citep{zimu:JWST}. Each image depicts one supra-root cell and a 5" scale bar is added on the top-left corner for reference. Most massive: left column. Second most massive: right column. Top row: dust free. Center row: no dust sublimation model. Bottom row: instant dust sublimation model.}
\label{fig:UVC}
\end{figure*}

We also produce false-color visual representations of UV starlight in this band of the two most massive galaxies in the snapshot of our C realization at $z\sim6$ for all three cases of dust implementation, Gaussian smoothed to twice the pixel scale of the short-wavelength FPAs of NIRCam on JWST, at 0.064" \citep{zimu:JWST} (figure \ref{fig:UVC}). RGB colors are generated by dividing our spectral range specified previously into three equal segments and applying the scaling from \citep{zimu:RGB} with $$F(x)=\sqrt{\sinh^{-1}{(x/x_c)}}$$ stretch, where $x_c=\bar{x}/8$ is an arbitrary choice that appears to maximally enhance galactic structures. This gives a better color resolution than that of NIRCam on JWST which has only two bands covering the relevant spectral range \citep{zimu:JWST}. Nevertheless, figure \ref{fig:UVC} shows that a substantial amount of details on galactic structures can be revealed by JWST even at $z\sim6$. The center row shows dust without sublimation (a simplest model of the dust-to-gas ratio being proportional to metallicity). Here dust is concentrated in regions which have active star formation or have had such in the past, highly reddening the entire galaxies except the outermost regions. The crucial importance of accounting for dust sublimation and spallation is shown in the bottom row, that presents the instant sublimation model - in that model the overall amount of reddening is substantially reduced. JWST therefore has the potential to put constraints on the complex dynamics of dust creation and destruction in the ISM at high redshifts directly through observing spatial distributions of dust in the galactic ISM. Another interesting feature of dust in the ISM revealed by figure \ref{fig:UVC} is elastic scattering of UV photons by dust grains. This means that dust glows even in UV. The bottom row shows that some regions with essentially no stars are illuminated once dust is introduced. This effect is actually more pronounced with dust sublimation, as there are no stars in these regions to destroy dust grains, but sublimation elsewhere allows more UV photons to elastically scatter on these dust grains. It is possible that such glow of dust in UV can be observed from deep fields with JWST; and comparing them to the glow of dust in IR due to re-emission can put further constraints on the compositions of dust grains.

\subsubsection{UV Luminosity Functions}

\begin{figure}[htbp]
\includegraphics[width=1\hsize]{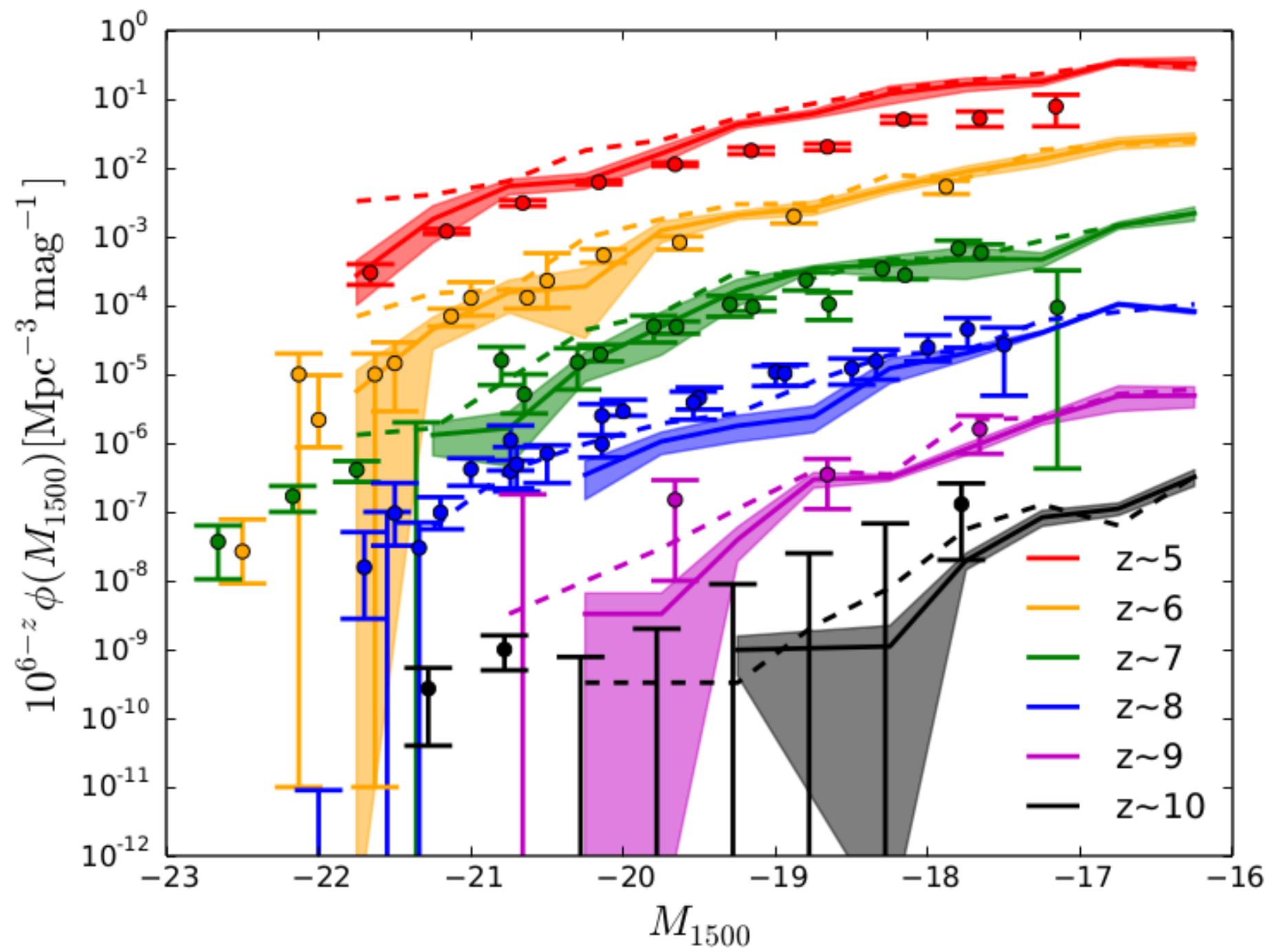}
\includegraphics[width=1\hsize]{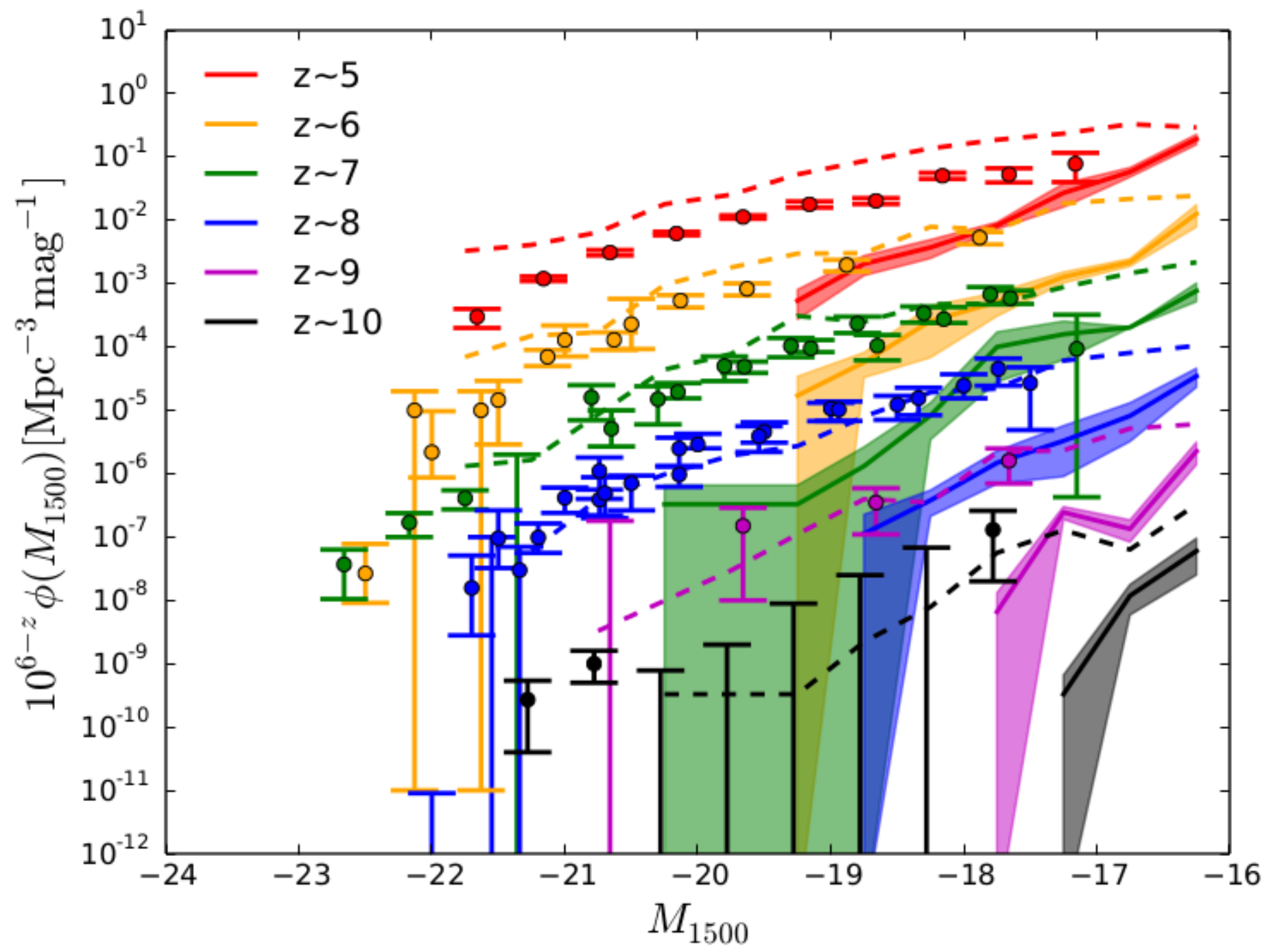}
\caption{Ultraviolet galaxy luminosity functions for our combined simulation sets displayed from low $z$ to high $z$ in the rainbow order, with black representing $z\sim10$ (average with dust - solid lines, standard errors - shaded, average without dust - dashed lines, standard errors are suppressed in the last case for clarity). Observations from \citet{gals:biff07,gals:biol11,rei:obig12,rei:btos12,rei:sreo13,rei:wmhb13,rei:obil13,rei:obi14,rei:bdm14} are represented by data points with error bars. Each $z$ is shifted vertically by 1 dex for clarity. Top: instant sublimation dust model. Bottom: no sublimation dust model.}
\label{fig:lumfunc}
\end{figure}
Figure \ref{fig:lumfunc} presents the evolution of UV luminosity functions of galaxies in our simulation sets from $z\sim10$ to $z\sim5$. Compared to the earlier work on CROC \citep{ng:g14}, our luminosity functions are more scattered, do not extend as far at the bright end, and often have larger standard errors, despite utilizing simulation sets with similar combined volumes. This is a direct consequence of sub-sampling galaxies in the present work due to the limitation of computational resources, as our Monte-Carlo radiative transfer calculations are much more computationally expensive than the simplistic dust model employed in \citet{ng:g14}. Also included is a compilation of recent observations of the UV luminosity function at these redshifts \citep{gals:biff07,gals:biol11,rei:obig12,rei:btos12,rei:sreo13,rei:wmhb13,rei:obil13,rei:obi14,rei:bdm14} marked by data points with error bars.

In both top and bottom panels, UV luminosity functions of the original starlight without dust obscuration are shown. They are in agreement with observations down to $z\sim8$, and begin overestimating the data from $z\sim7$ to lower redshifts. This result is consistent with current observations, which suggest negligible dust obscuration at $z\sim8$ and above \citep{gals:bifc09,gals:biol11}.

Theoretical models of galactic dust evolution \citep{ism:i03,ism:i11,zimu:Asano2013,zimu:Feldmann2015} suggest non-linear scaling between the dust-to-gas ratio and metallicity for metal-poor galaxies, predicting lower dust-to-metal ratios, and therefore further reducing the amount of dust obscuration at high redshifts. Given known potential candidates for dust grains in the ISM \citep{ism:d09,misc:wd01}, there is little reason to believe that the extremely low UV obscuration as suggested by the comparison between the observed and our simulated UV luminosity functions is caused by some intrinsic differences in dust grains at high redshifts that dramatically alter their UV optical properties. Thus our UV luminosity functions indicate very low dust abundance at $z\gtrsim8$ as predicted by theoretical models.

UV luminosity functions of the original starlight for $z<8$ are systematically above observations for bright galaxies ($M_{1500}\lesssim-19$). It may appear that there is also overestimation for faint galaxies at $z\sim5$. Even though this is true as far as comparisons with data are concerned, we note that for these galaxies the dust free UV luminosity function completely overlaps with that with instantly sublimated dust. This is also true for faint galaxies at higher redshifts, indicative of low dust abundance in these galaxies. Thus, the overestimation for faint galaxies at $z\sim5$ is not due to the UV luminosity function being dust free, but instead is due to the intrinsic limitation of our simulations. The dust free UV luminosity functions unambiguously deviate from those with instantly sublimated dust for bright galaxies, and therefore the overestimation for these galaxies is clearly caused by the unrealistic lack of dust in the ISM where the dust free UV luminosity functions are computed. This is clear evidence that there is a sufficient quantity of dust existing in the ISM of bright model galaxies at $5\lesssim z<8$ to cause an amount of UV obscuration already detectable by current observations.

On the other hand, the presence of dust in faint galaxies remains below current detection limits for all redshifts analyzed. Note, that except for one data point at $z\sim7$, error bars in the data do not enlarge significantly for fainter galaxies, thanks to enhanced statistics. Thus, the lack of overestimation of the dust free UV luminosity functions for these galaxies at $z<8$ (excluding that due to factors caused by intrinsic limitations of the simulations themselves) indicates that dust content in the ISM is dependent on galaxy UV luminosity. 

As described in subsection \ref{subsec:dust}, a commonly adopted treatment of cosmic dust in numerical simulations is to assume a linear scaling between the dust-to-gas ratio and metallicity, a reasonable assumption for a quasi-static equilibrium state given the increasingly clear evidence of ISM accretion being the dominant mechanism for dust growth \citep{ism:d09,ism:i11,ism:i03,zimu:Asano2013,zimu:Feldmann2015}. However, the bottom panel of figure \ref{fig:lumfunc} unambiguously disproves this treatment in our redshift range. These UV luminosity functions are results of the canonical assumption without any dust sublimation, and they significantly overestimate dust contents of galaxies regardless of UV luminosity. Of particular interest is the fact that the slopes of these UV luminosity functions are somewhat steeper than those suggested by observation, a consequence of the generally higher metallicity in more massive galaxies. It also means that the discrepancy cannot be eliminated through decreasing the assumed dust-to-metal ratio. The discrepancy also appears to widen with increasing redshift. Because of such dramatic failure, we do not consider the no-sublimation model hereafter.

The instant dust sublimation model outlined in subsection \ref{subsec:dust} is shown in the top panel. The luminosity dependent downward corrections from dust free UV luminosity functions are more significant at the bright end, indicating higher dust contents in the ISM of more massive galaxies. The dust correction brings these UV luminosity functions into better agreements with observations. Complete agreements at all luminosities within current observational limits are found at $z\sim6-7$. The overestimation for faint galaxies at $z\sim5$ results from the intrinsic limitation of our simulations mentioned above. We attribute this primarily to degradation of spatial resolution at this redshift described in section \ref{sec:sims}, which is relatively more severe for smaller galaxies. As we have seen, galaxies at $z\gtrsim8$ are consistent with being dust-free. Our instant dust sublimation model is unable to sublimate away a sufficient amount of dust for brighter galaxies in this redshift range, resulting in underestimations at the bright end; the data at $z\gtrsim9$ is too sparse to constraint dust models, but the $z\approx8$ measurements are clearly inconsistent with the instant dust sublimation model and require essentially dust-free ISM.

\subsubsection{UV Continuum Slopes}

\begin{figure*}[htbp]
\includegraphics[width=\textwidth]{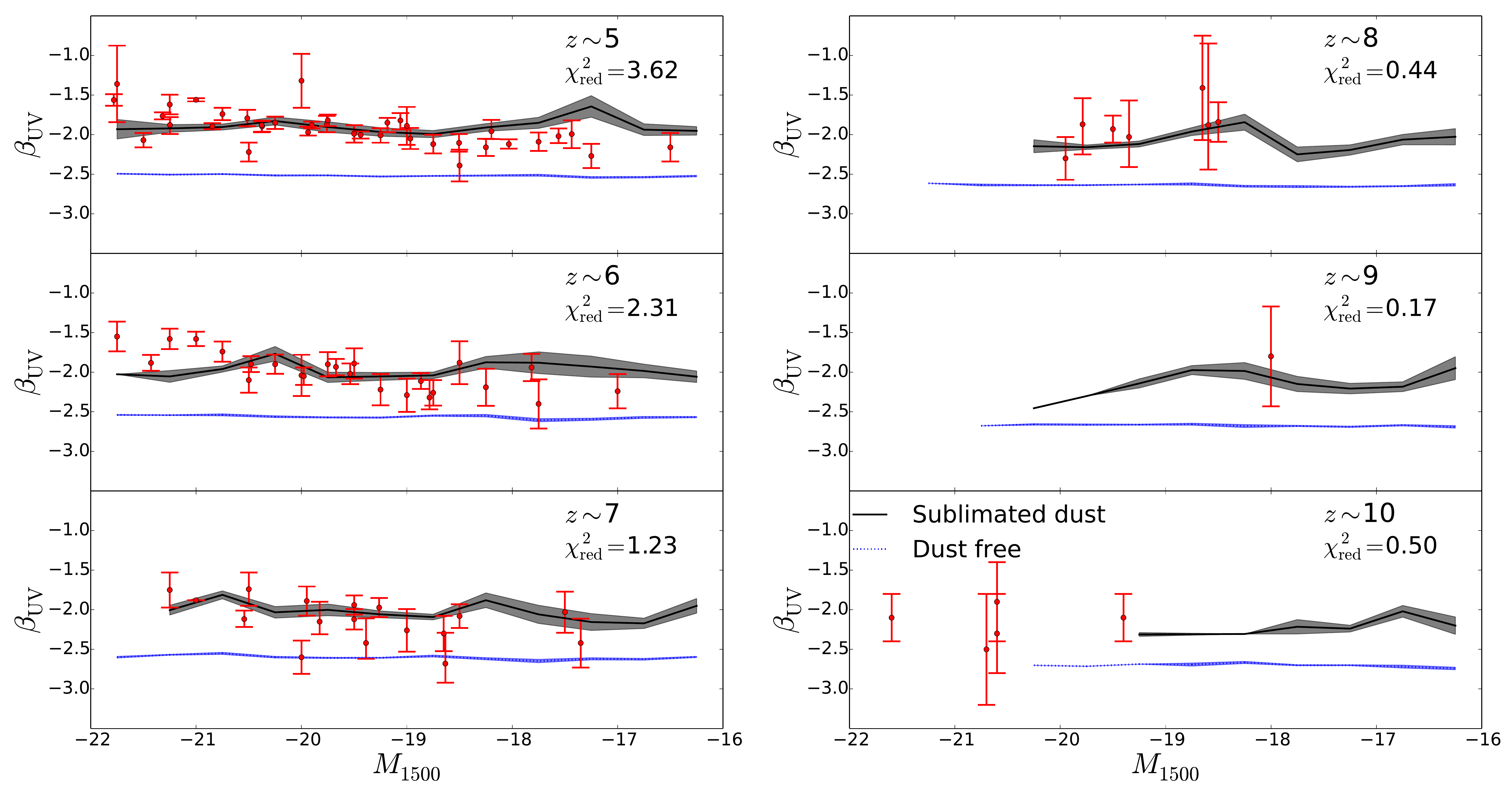}
\caption{UV continuum slopes of SEDs for galaxies in our simulation sets averaged over UV magnitude bins. Black solid lines represent UV slopes of dust with instant sublimation, and blue dotted lines show UV slopes without dust. Bands represent standard error of the mean for each magnitude bin. Data points with error bars showing standard errors are a compilation of observations \citep{rei:biol14,rei:rmd14,zimu:Duncan2014,zimu:Dunlop2012,zimu:Dunlop2013,zimu:Finkelstein2012,zimu:Wilkins2011}.}
\label{fig:beta}
\end{figure*}

The UV luminosity functions shown in figure \ref{fig:lumfunc} encode information on dust obscuration at the representative wavelength of $1500\angstrom$ in the rest frame. However dust obscuration is wavelength dependent, and is higher at shorter wavelengths for the wavelength range analyzed \citep{ism:d09,misc:wd01}. Thus, the presence of dust grains in the ISM produces a reddening effect on UV starlight, typically quantified by fitting a power-law to the spectral flux density,
\begin{equation}
F_{\lambda}\propto\lambda^{\beta_{\mathrm{UV}}},
\label{eq:beta}
\end{equation} where the UV continuum slope, $\beta_{\mathrm{UV}}$, characterizes the amount of reddening.

Figure \ref{fig:beta} shows $\beta_{\mathrm{UV}}$ as a function of UV luminosity for galaxies in our simulations at $z\sim5 - 10$, both with dust obscuration and for the original starlight only, computed by averaging $\beta_{\mathrm{UV}}$ for galaxies over UV magnitude bins of 0.5 magnitudes. For the case of the original starlight, our star formation model produces flat UV continuum slopes at $\beta_{\mathrm{UV}}\sim -2.5$ at $z=5$  that decrease slightly with increasing redshift as younger, bluer stars are more dominant in the earlier universe. The UV continuum slopes with dust obscuration exhibit more significant fluctuations over UV magnitude bins, probably partly caused by lack of statistics as in Figure \ref{fig:lumfunc}. However we note that some features of these fluctuations do appear to be corroborated by the compilation of observations  \citep{rei:biol14,rei:rmd14,zimu:Duncan2014,zimu:Dunlop2012,zimu:Dunlop2013,zimu:Finkelstein2012,zimu:Wilkins2011}. In all cases, UV continuum slopes with dust computed from our simulations are in relatively good agreements with observations, while those of the unobscured starlight clearly do not agree with the measurements, indicating a detectable amount of dust reddening at all redshifts where observations currently exist. This is a significant improvement from the earlier work on CROC \citep{ng:g14} where agreement was only obtained at $z\sim5$ and only with a subset of observations. The improvement serves as confirmation that our instantly sublimated dust coupled with Monte Carlo radiative transfer is indeed more realistic than the simplistic dust obscuration model employed in the earlier work. For dust without sublimation, we have $-1.0\lesssim\beta_{\mathrm{UV}}\lesssim0.5$ including all redshifts analyzed, much too red for any kind of agreement with observations. This is another piece of evidence for the necessity of accounting for the complex dynamics of dust creation and destruction in the ISM, and that a naive application of the canonical assumption can be far from reality. 

Redshift evolution confirmed by observations of UV luminosity functions is indeed also followed by $\beta_{\mathrm{UV}}$, which decreases overall with increasing redshift by amounts greater than those due to bluer stars being more dominant at higher redshifts, indicating lower dust contents in the ISM of galaxies at higher redshifts. The sizes of the error bars in observational data prevent one from drawing conclusions on the redshift evoluton of $\beta_{\mathrm{UV}}$ in the actual data. But the hints given by the data do appear to be in agreement with our simulated results.

Even though the UV continuum slopes from our simulations are completely independent of direct observations, and are not fits to the data, we compute the reduced $\chi^2$ value at each redshift where data exist as a zero-parameter fit taking into account both observational and theoretical uncertainties to quantify the agreement of our results with data. It is not entirely surprising that $\chi^2_{\mathrm{red}}$ decreases with increasing redshift given more uncertain data at earlier universe. The $\chi^2_{\mathrm{red}}$ values at $z\sim5$ and $z\sim6$, while may possibly be improved somewhat through reducing fluctuations by increasing statistics, do seem to indicate that although our current treatment of dust in the ISM is a significant improvement from that of the earlier work, it is still a rather crude approximation to the complete physical processes that govern dust formation, evolution, and destruction in the ISM.

\begin{figure*}[ht]
\includegraphics[width=\textwidth]{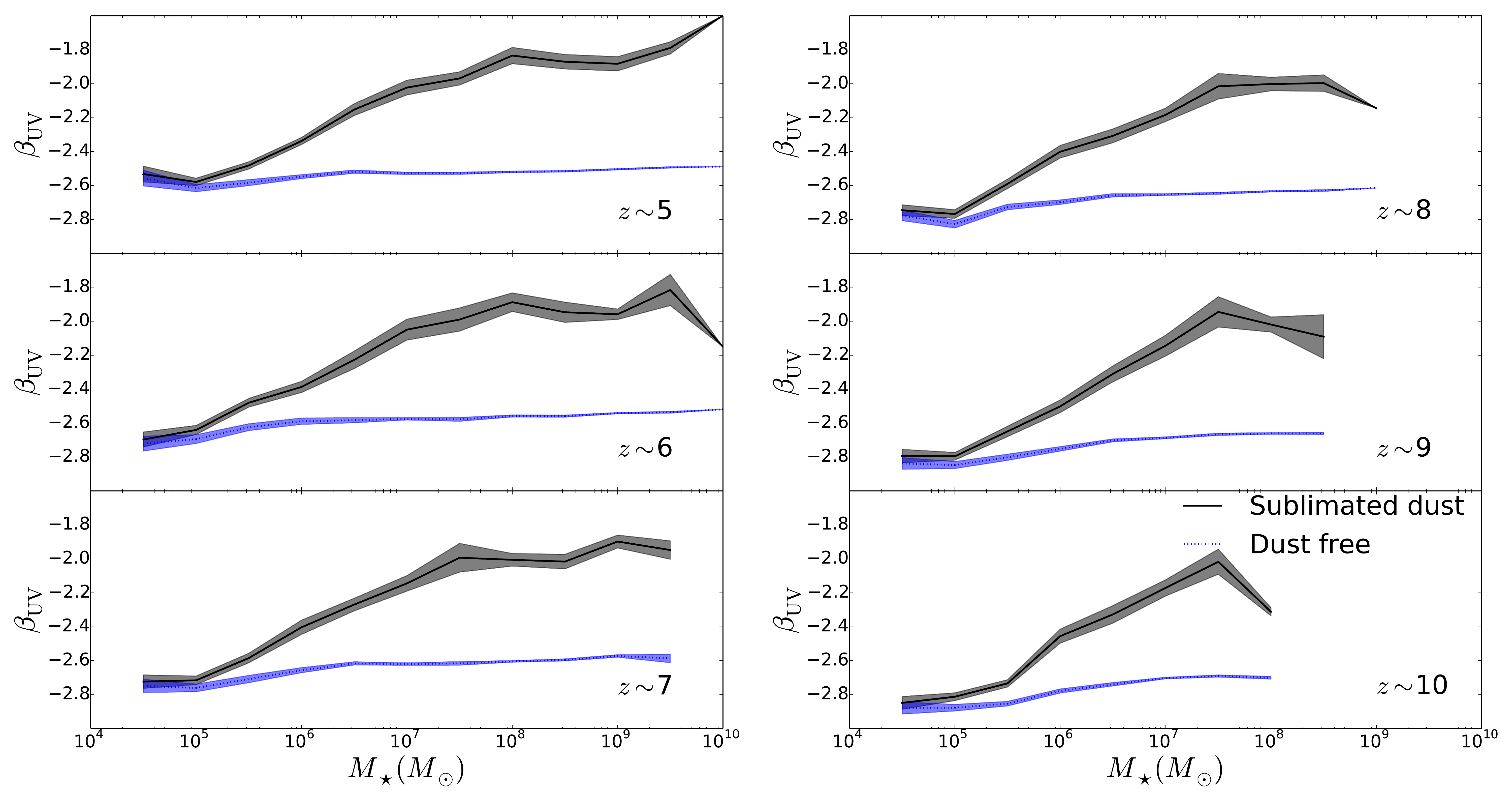}
\caption{UV continuum slopes of SEDs for galaxies in our simulation sets averaged over stellar mass bins. Black solid lines represent UV slopes of dust with instant sublimation, and blue dotted lines show UV slopes without dust. Bands represent standard error of the mean for each stellar mass bin.}
\label{fig:betam}
\end{figure*}

As elucidated previously, we expect from theoretical treatments higher dust contents in the ISM of larger galaxies. Observed UV luminosity functions in figure \ref{fig:lumfunc} appear to confirm this expectation. Thus one would similarly expect more significant reddening of larger galaxies, a feature completely absent from our results in figure \ref{fig:beta}. Any potential trends in the data are also rather weak. This result also stands somewhat in contrast with that from the earlier work where a slight rising trend of $\beta_{\mathrm{UV}}$ with UV luminosity appears to exist. Moreover, the inconsistency between starlight $\beta_{\mathrm{UV}}$ and data at all redshifts where observations currently exist seems to present some tension with the observations of UV luminosity functions which indicate undetectable amounts of dust obscuration in smaller galaxies. One factor neglected by the earlier work is the complicated dust geometry from distributions of metals and HII regions in the galaxies. This could play a role in altering the UV luminosity dependence of $\beta_{\mathrm{UV}}$. Elastic scattering of UV photons by dust particles described in subsection \ref{subsec:radtrans} is another factor neglected in the simple model of the previous work. However this factor has an opposite effect on figure \ref{fig:beta} to what is needed to explain the flatness. Figure \ref{fig:N} shows that elastic scatterings due to dust albedo reduce the effective column density at longer wavelengths, as dust albedo increases quite sharply with wavelength in our spectral range \citep{ism:d09,misc:wd01}, further lowering the amount of absorption of redder light. Thus elastic scatterings enhance reddening due to dust obscuration, and therefore increase $\beta_{\mathrm{UV}}$ by a greater amount for galaxies with higher dust contents. We have verified this fact by setting dust albedo to 0 at all wavelengths, which makes galaxies with higher UV luminosities even bluer, skewing the trend in figure \ref{fig:beta} to an even more perplexing direction. 

Another possible cause of the flatness of $\beta_{\mathrm{UV}}$ as a function of galaxy luminosity is the mixing of galaxies with various stellar masses undergoing starbursts at different points in time, which induces dramatic changes in their UV luminosities. Then, on average, the galaxies would be constantly shifted around in UV magnitude bins, and the resultant mixing would cause differences in $\beta_{\mathrm{UV}}$ to average out for each UV magnitude bin, flattening the curves in figure \ref{fig:beta}. We have tested this possibility and found that UV luminosities correlate well enough with stellar masses in our model galaxies to make the effect unimportant.

To explore the discrepancy with the observational trends further, we show in Figure \ref{fig:betam} UV slopes vs stellar masses for our model galaxies over the full range of galactic masses. Indeed, there is a strong correlation between the stellar masses and average UV slopes, but it saturates for $M_*\la10^8{\rm M}_\odot$, within the observed mass range, because all these galaxies have similar gas-phase metallicites.

\subsubsection{UV Optical Depths}

As discussed in previous sections, the presence of dust in galaxies can be detected through its reddening effect due to wavelength-dependent effective UV extinction by dust grains, as well as through observing UV luminosity functions which serve as indirect measurements of UV extinction due to dust at a particular representative wavelength. In principle, results from the latter provide an overall normalization on results from the former, and therefore combining them together can produce a relatively complete characterization of UV absorption by dust grains, from which one can extract information about the grains themselves. However, obtaining UV extinction from UV luminosity functions requires knowledge of UV luminosity functions without dust obscuration, and is therefore dependent on stellar populations of the galaxies as well as modelling of star formation histories. Thus it is desirable to obtain UV extinction by dust grains from observed quantities without relying on excessive modelling.

One widely adopted method to measure UV extinction relatively independently of star formation models is to rely on the reddening effect of dust to translate observed values for the UV continuum slope $\beta_{\mathrm{UV}}$ into the effective dust opacity \citep[c.f.][]{gals:bifc09}. A commonly used calibration for such a measurement is due to \citet{zimu:MHC}, who obtained a linear correlation between UV extinction $A_{1500}$ and UV continuum slope $\beta_{\mathrm{UV}}$,
\begin{equation}
\label{eq:A1500}
A_{1500}=1.99(\beta_{\mathrm{UV}}-\beta_0),
\end{equation}
where $\beta_0=-2.23$ is the starlight UV continuum slope from the samples of \citet{zimu:MHC}. We adopt \unit[1500]{\angstrom} to be the representative wavelength throughout this study for consistency with our earlier study \citep{ng:g14}. Given the relatively slow variation of the dust opacity with wavelength in this range, this change from \citet{zimu:MHC} is of little importance. We can then derive the UV optical depth
\begin{equation}
\label{eq:tau}
\tau_{1500}\equiv -\ln\left(\frac{F_{\mathrm{d}}}{F_0}\right)=\frac{A_{1500}}{1.086},
\end{equation}
where $F_{\mathrm{d}}$ is the spectral flux density of a galaxy with dust at the representative wavelength obtained by performing radiative transfer, and $F_0$ is that of the original starlight at the same wavelength. This empirical conversion simply states that dust UV optical depth is linearly proportional to the amount of dust reddening, an intuitive correlation as more dust reddening probably indicates a higher dust content and therefore obscuration.

\begin{figure}[t]
\includegraphics[width=1\hsize]{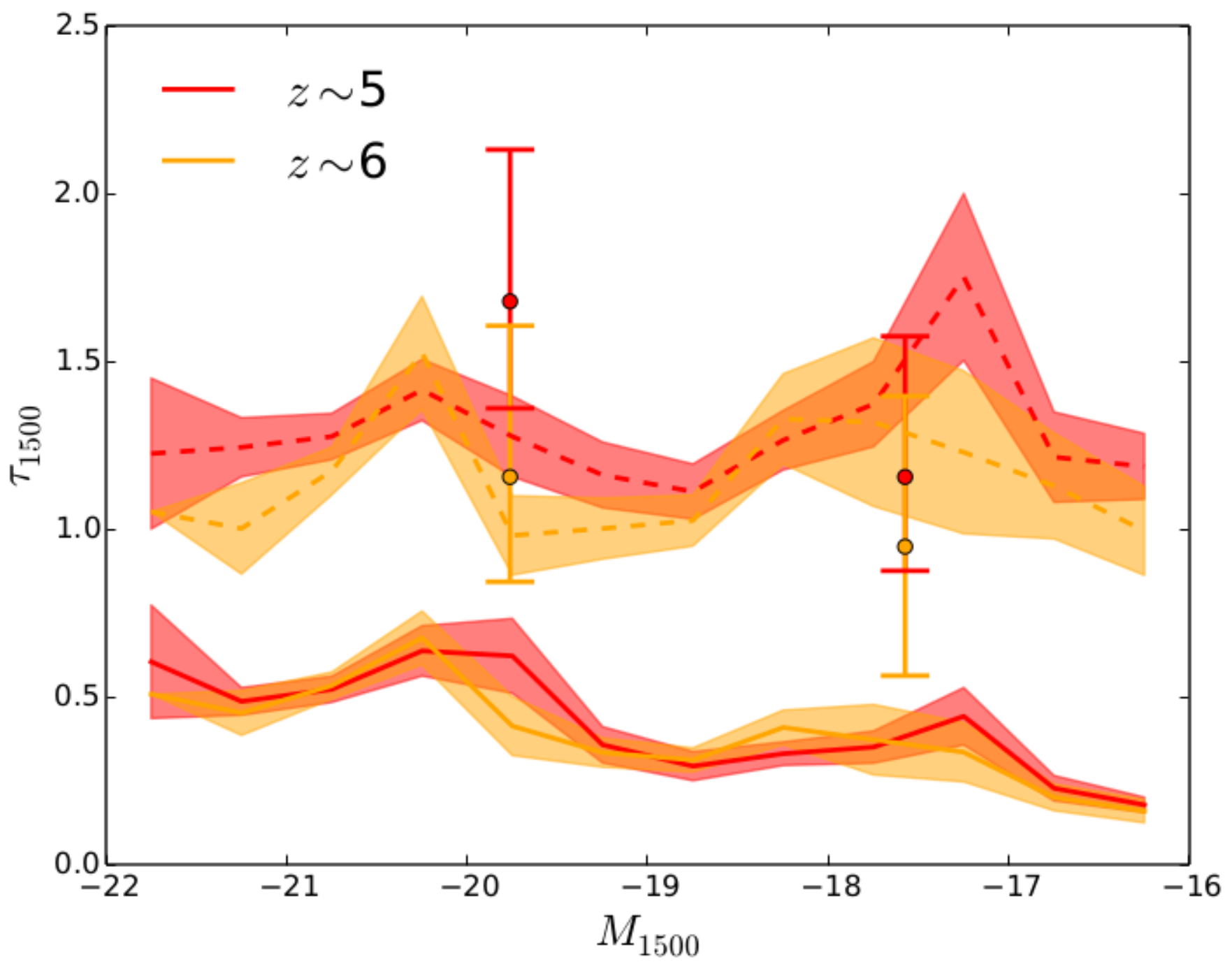}
\includegraphics[width=1\hsize]{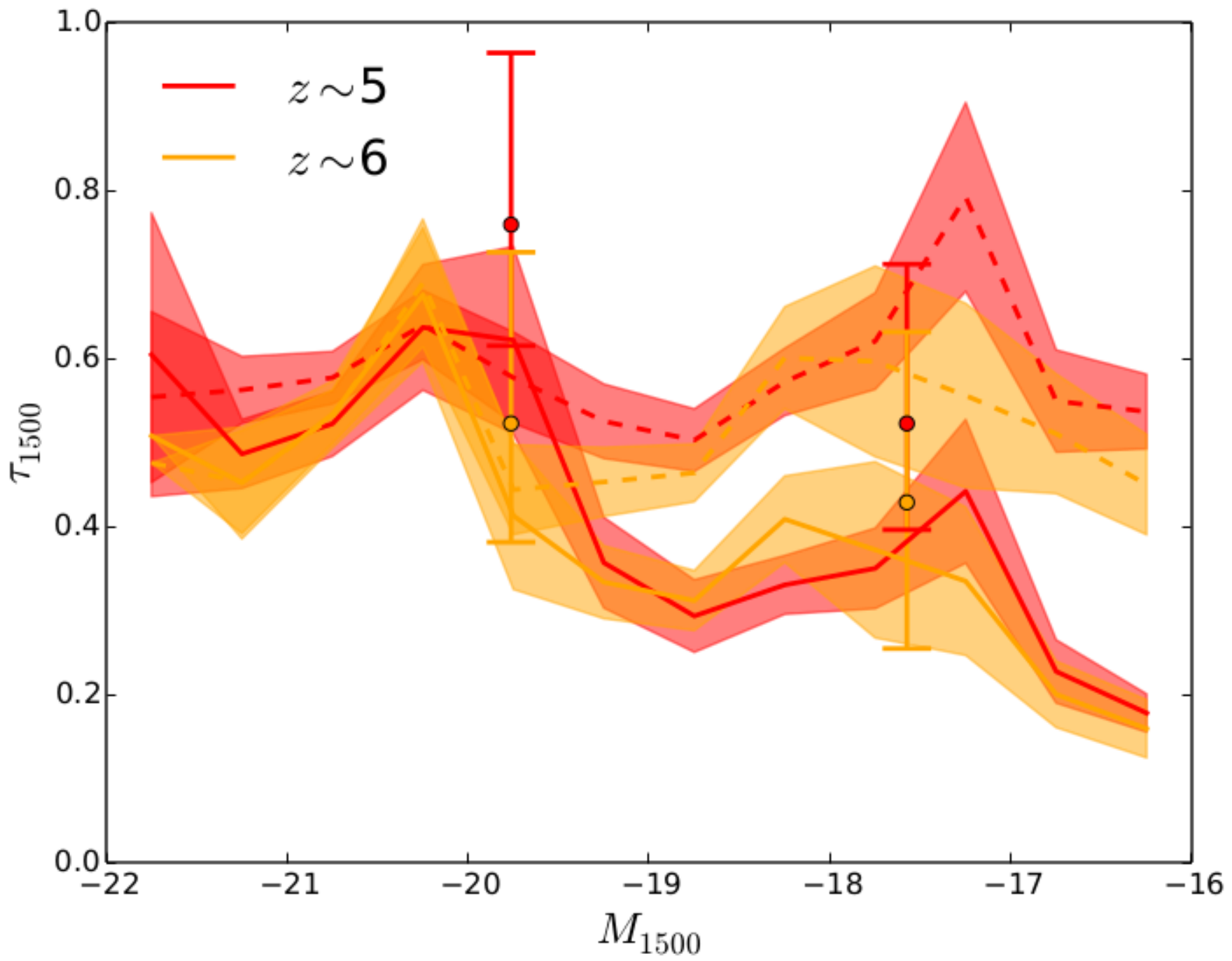}
\caption{Ultraviolet optical depth of dust in galaxies in our simulation sets averaged over ultraviolet magnitude bins at $z\sim 5$ and 6 (direct computation - solid lines, $\beta_{\mathrm{UV}}$ to $A_{1500}$ conversion - dashed lines). Shades illustrate $1\sigma$ standard errors of the mean. Data points with error bars are observations from \citet{gals:bifc09} shifted to $\beta_0=-2.6$. Top: the slope of the $A_{1500}$ - $\beta_{\mathrm{UV}}$ correlation from \citet{zimu:MHC}. Bottom: the slope of the $A_{1500}$ - $\beta_{\mathrm{UV}}$ correlation decreased to $0.9$.}
\label{fig:tau}
\end{figure}

Though it is not yet possible to perform the necessary dust reemission measurements to verify equation (\ref{eq:A1500}) for high-redshift galaxies, this correlation is nevertheless often adopted for observations of such galaxies. Since we are able to directly compute $\tau_{1500}$ from definition for galaxies in our simulation sets, we are afforded an opportunity to verify the $\beta_{\mathrm{UV}}$ - $\tau_{1500}$ conversion using equations (\ref{eq:A1500}) and (\ref{eq:tau}) by performing a comparison between our results and high-redshift observations of $\tau_{1500}$. The solid lines in figure \ref{fig:tau} represent actual $\tau_{1500}$ computed directly from our simulation sets at $z\sim 5$ and 6. These lines are in good agreement with values reported by \citet{gals:bifc09}. However it would be incorrect for one to simply use these results from \citet{gals:bifc09}, as our unobscured stellar populations at these redshifts are bluer ($\beta_0\sim -2.6$) than what is assumed by \citet{zimu:MHC} ($\beta_0=-2.23$). The effect of the intrinsic starlight color is even more pronounced in section \ref{sec:reemit}, where $\beta_0=-2.23$ produces dramatic disagreement. Adopting $\beta_0\sim -2.6$ results in a shift of $\Delta\tau_{1500}=0.68$ from measurements of \citet{gals:bifc09}. Now observations are clearly not in agreement with our $\tau_{1500}$ within $1\sigma$. To test if the disagreement is due to intrinsic differences in dust properties or simply an artifact of the conversion, we perform a similar conversion on $\beta_{\mathrm{UV}}$ obtained from our simulation sets with $\beta_0\sim -2.6$, and the results are illustrated by dashed lines in the top panel of figure \ref{fig:tau}. Indeed our simulated results computed through the $\beta_{\mathrm{UV}}$ - $\tau_{1500}$ conversion are essentially in agreement with observations from \citet{gals:bifc09} shifted to account for $\beta_0$. This indicates that most likely the disagreement between the actual $\tau_{1500}$ computed from our simulation sets and observations from \citet{gals:bifc09} is due to the assumed $\beta_{\mathrm{UV}}$ - $\tau_{1500}$ conversion.

We further endeavor to modify the correlation (\ref{eq:A1500}) in an effort to obtain better agreements between solid and dashed lines in figure \ref{fig:tau}, as well as with observations, which would indicate a more realistic conversion relation. With $\beta_0$ being a physical parameter of our stellar populations, the only free variable is the slope set to be 1.99 by \citet{zimu:MHC}. The bottom panel of Fig.\ \ref{fig:tau} shows the comparison between the ``true'' values of $\tau_{1500}$ with the values obtained from (\ref{eq:A1500}) with the numerical coefficient in front of $(\beta_{\mathrm{UV}}-\beta_0)$ changed from 1.99 to 0.9. With the normalization of the conversion relation thus adjusted, we find an agreement for the brightest galaxies, which, however, breaks down at $M_{1500}>-19.5$. The shapes of solid and dashed lines in Fig.\ \ref{fig:tau} are too different for \emph{any} value for the normalization coefficient to work. More than that, since the dependence of $\beta_{\mathrm{UV}}$ on galaxy magnitude is weak, even a more complex, non-linear conversion from $\beta_{\mathrm{UV}}$ to $\tau_{1500}$ would not be able to reproduce the actual dust opacities in our model galaxies. Hence, we conclude that for our model galaxies it is not possible to recover the actual dust opacity $\tau_{1500}$ from the UV continuum slope $\beta_{\mathrm{UV}}$ at all.

\subsection{Infrared Reemission}
\label{sec:reemit}

Dust particles are heated by energy gained from absorbing UV photons from starlight. The temperatures of dust particles are in the range such that most of the thermal reemission of dust particles are in the far infrared spectral window. Dust IR reemission gives rise to a third way of observing the effects of dust, in addition to UV absorption and scattering. At $z\sim6$ the peak of dust reemission falls around $400-500\mu\mathrm{m}$, well within the spectral range of ALMA. While the spatial resolution of ALMA is not sufficient to resolve galaxies at $z\la 6$, it is possible to measure IR fluxes and, possibly, CII and CO emission lines from these galaxies.

Dust IR reemission is often quantified with the so-called ``infrared access'' (IRX), which is defined as a ratio of far IR (FIR) flux in the band $\unit[42.5]{\mu m}$ - $\unit[122.5]{\mu m}$ in the rest frame to UV flux \citep{zimu:MHC},
\begin{equation}
\mathrm{IRX}_{1500}=\frac{\mathrm{FIR}}{F_{1500}}.
\end{equation}

\begin{figure}[t]
\includegraphics[width=1\hsize]{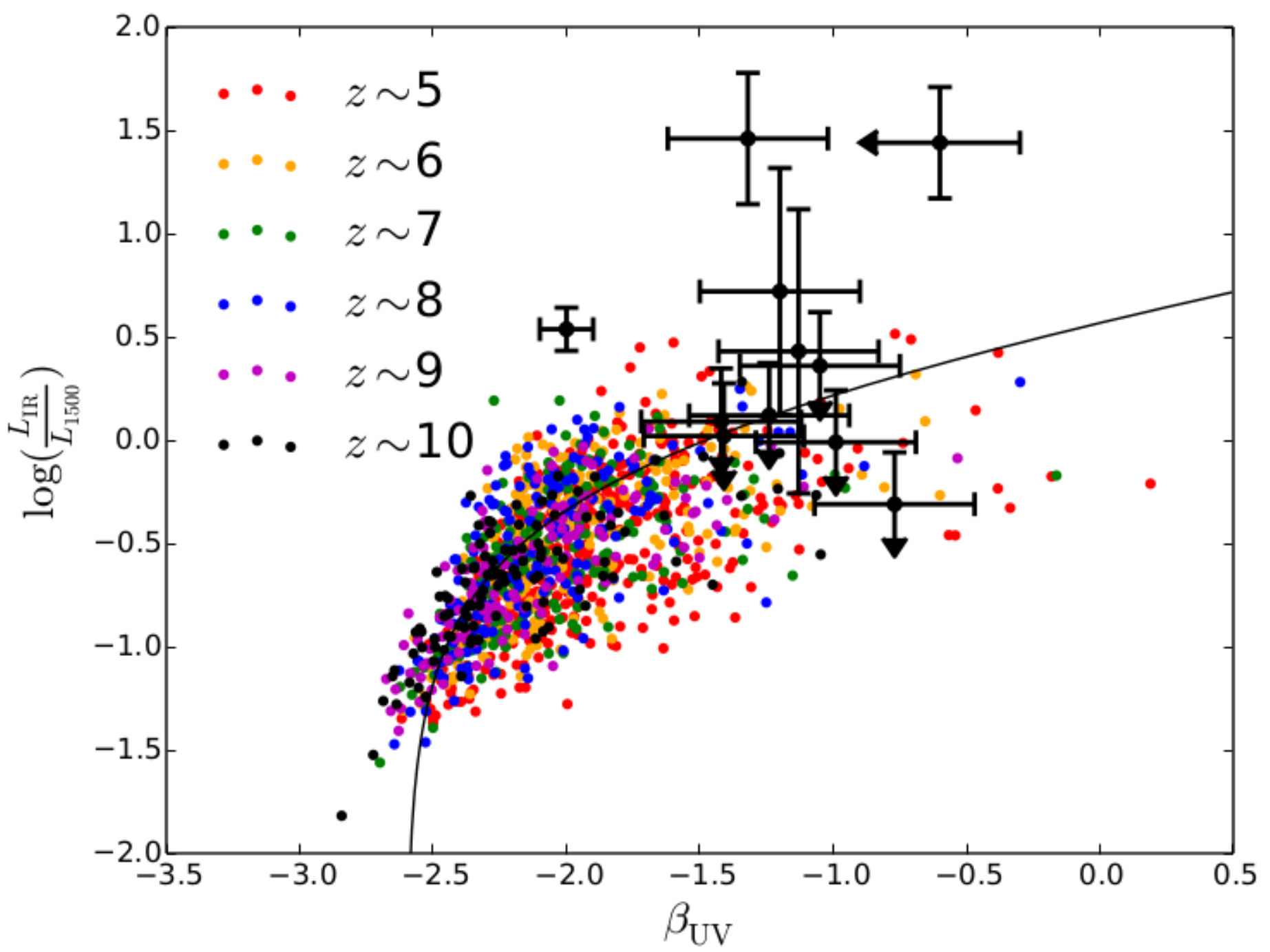}
\caption{Infrared excess from dust reemission in the instant sublimation model as a function of the ultraviolet continuum slope for simulated galaxies at several redshifts, displayed from $z=5$ to $z=10$ in the rainbow order. Contribution from Ly-$\alpha$ resonant scatterings \citep{zimu:lalpha} is included. The solid line is a fit from \citet{zimu:MHC} with $\beta_0=-2.6$. Observations from \citet{zimu:watson15} and \citet{zimu:capak15} are represented by data points with error bars.\label{fig:irxb}}
\end{figure}

Figure \ref{fig:irxb} shows IRX versus the UV continuum slope for our simulated galaxies, and compares them to the recent observational measurements and upper limits \citep{zimu:watson15,zimu:capak15}. IRX is computed from our simulations by heating dust particles with entire spectra of stellar particles as described in section \ref{sec:rad}. However, an issue with correctly computing FIR and therefore IRX is to properly account for the not insignificant energy contribution from Ly-$\alpha$ resonant scatterings, which cannot be included in stellar spectra, due to both spectral resolution and the fact that the UV optical depth near Ly-$\alpha$ is enormous, rendering normal radiative transfer computations infeasible. But for the same reason essentially all of the energy close to this wavelength is absorbed by dust, and therefore contribution to FIR from this process has to be specifically included. We adopt the estimate of \citet{zimu:lalpha} which puts the energy contribution from Ly-$\alpha$ photons to be $\sim 7\%$ of the rest of UV contribution. Just like with UV slopes, we find reasonable general agreement here, although our simulation boxes are barely large enough to include enough observable galaxies (the overlap range will widen when the data from the full ALMA become available). The curve in figure \ref{fig:irxb} is the fit adopted by \citet{zimu:MHC} with $\beta_0$, the dust-free UV continuum slope, shifted from $-2.23$ to $-2.6$ to be suitable for our adopted  stellar population synthesis models. The fact that most of our simulated galaxies lie close to the curve is expected, as the model derived by \citet{zimu:MHC} is nothing more than a statement of conservation of energy. Nevertheless, this agreement serves as a simple sanity check. The lack of simulated galaxies with IRX above 0.5 may be due to the small sizes of simulated and observational samples, or may be due to our assumed dust ansatz; at present, it is too early to say.

\section{Conclusions}
\label{sec:concl}

By post-processing simulated CROC galaxies with radiative transfer in the dusty medium with the \textsc{Hyperion} code, we are able to make theoretical predictions for a large set of currently observable characteristics of high redshift galaxies, such as their UV luminosities, UV continuum slopes, and FIR luminosities. Since CROC simulations do not model dust from the first principles yet, we need to assume an ansatz for the dust abundance. The simplest possible ansatz, dust follows metals everywhere, fails by a large margin (Fig.\ \ref{fig:lumfunc}, bottom panel). A more realistic ansatz, dust follows metals in the neutral gas (which we call ``instant sublimation model'') fares much better, providing reasonable overall agreement with all of the existing observational data.

Nevertheless, we also find intriguing discrepancies between theoretical expectations and actual data. First, it appears that simulations predict slightly too much dust at $z\sim8$, but not at earlier or later times. This discrepancy cannot be explained away by the incorrect metals production in the simulations - since simulations reproduce the whole evolution of the UV luminosity functions, they recover correct (within the observational uncertainties) star formation histories for all observable galaxies. Hence, one can expect that the metallicities of model galaxies are reproduced as well, and the apparent discrepancy with the observations is caused by our assumed dust-to-metal ratio, rather than by the wrong metallicity distribution. 

Admittedly, this discrepancy is marginal, but if it is confirmed with the future JWST data, it would imply that either the dust in $z\sim8$ galaxies is present near hot stars in HII regions and destroyed more efficiently, or it did not have enough time to form as efficiently as at later times. Contrasting this fact to galaxies at $z\lesssim7$ suggests some kind of transition occurring between $z\sim7$ and $z\sim8$, with dust either surviving for longer or forming more efficiently at later times. This fact appears to be consistent with semi-analytical treatments of dust evolution in the ISM that feature a "critical metallicity" above which dust growth in the ISM becomes much more efficient \citep{ism:i11,zimu:Asano2013,zimu:Feldmann2015}. More massive galaxies reach this critical metallicity by $z\sim7$ according to our results and smaller galaxies never reach it in the redshift range analyzed.

In addition, we find less dependence of the UV continuum slopes on galaxy UV luminosities and less variation in that relation. Coupled with $z\sim8$ discrepancy in the luminosity function discussed above, this may reflect the complex interplay of various dynamical dust creation and destruction processes mentioned in subsection \ref{subsec:dust}, and described in detail elsewhere \citep{ism:i03,ism:i11,zimu:Asano2013,zimu:Feldmann2015}. Optimally, these mechanisms should be fully incorporated in the simulations to physically account for the effects of dust in the ISM. Even before fully dynamical simulations of dust in the ISM become possible, relatively simple static models that represent a possible quasi-static equilibrium state achieved through balancing the various dynamical mechanisms may already be warranted given current observational limits \citep{zimu:Feldmann2015}.

This conclusion may be further strengthened by future ALMA detections of dust reemission in high redshift galaxies. Our current simulations appear to produce less scatter in the infrared excess at a given UV continuum slope; while this difference is not yet statistically significant, it may become stronger as observations with full ALMA build up the high redshift sample.

Finally, on a side, we note that actual dust opacities of our model galaxies \emph{cannot} be derived from the observed UV continuum slopes, even if we allow an arbitrary, non-linear relation between $\beta_{\mathrm{UV}}$ and $\tau_{1500}$. Hence, one has to either rely on FIR observations or come up with a novel approach to actually measure the dust contents of $z\gtrsim6$ galaxies.

\acknowledgements 
Radiative transfer simulations used in this work have been performed on the Joint Fermilab - KICP Supercomputing Cluster, supported by grants from Fermilab, Kavli Institute for Cosmological Physics, and the University of Chicago. Fermilab is operated by Fermi Research Alliance, LLC, under Contract No.~DE-AC02-07CH11359 with the United States Department of Energy. This work was also supported in part by the NSF grant AST-1211190. This work made extensive use of the NASA Astrophysics Data System and {\tt arXiv.org} preprint server.

\bibliographystyle{apj}

\bibliography{ng-bibs/rei,ng-bibs/newrei,ng-bibs/self,ng-bibs/sims,ng-bibs/misc,ng-bibs/gals,ng-bibs/ism,zimu}

\appendix
\section{Comparison between Color Conversions and Direct Fits}
\begin{figure}[htbp]
\centering
\includegraphics[width=\textwidth]{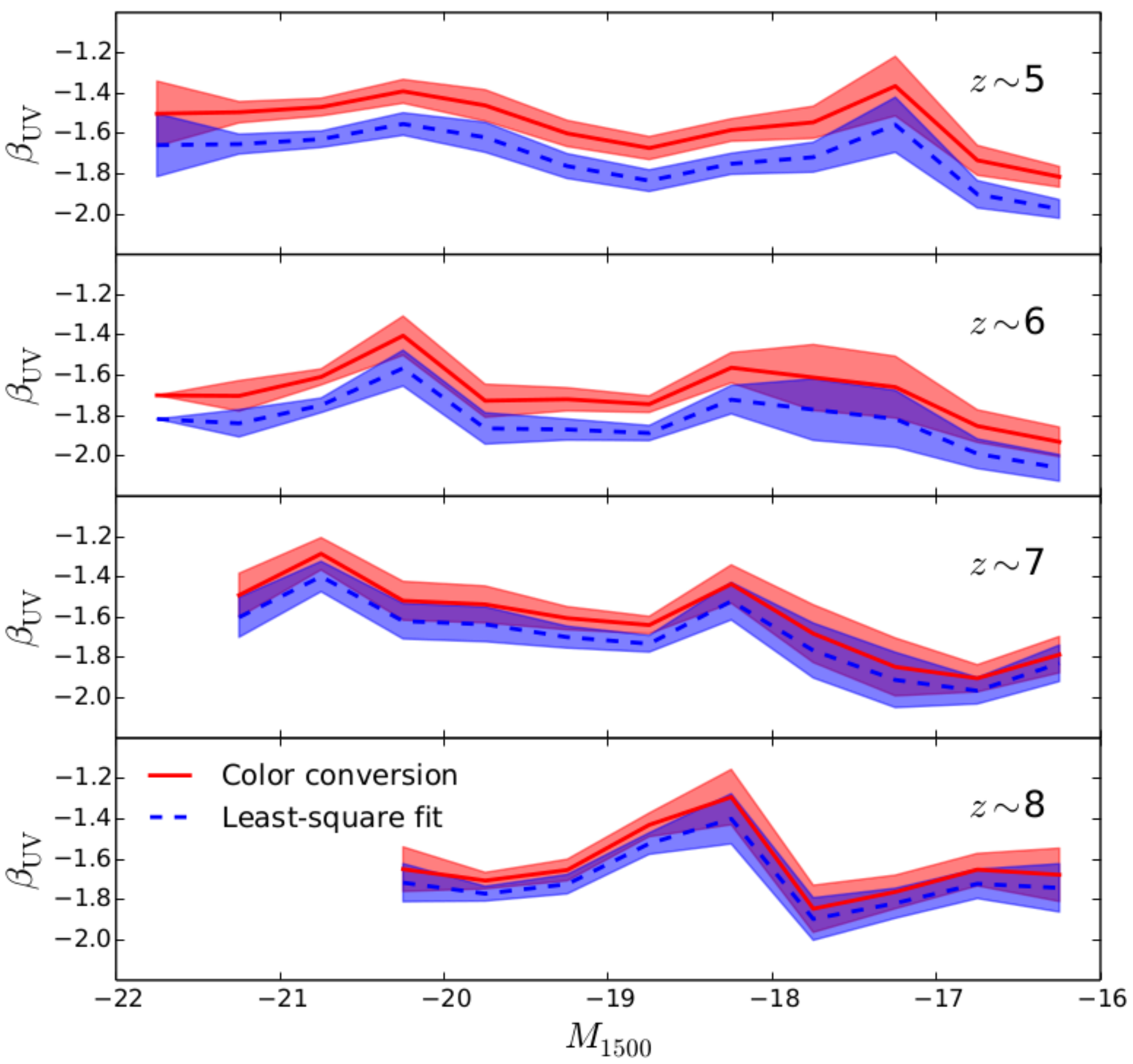}
\caption{UV continuum slopes of SEDs for galaxies in our simulation sets averaged over UV magnitude bins computed from color conversions \citep{rei:bio10,rei:biol14} (red solid lines) and direct least-square fits (blue dashed lines) with instant dust sublimation. Bands show the standard error of the mean for each magnitude bin.}
\label{fig:cc}
\end{figure}

Throughout our analyses we have utilized direct least-square fits for the flux versus wavelength to obtain UV continuum slopes as they follow simply from the definition of $\beta_{\mathrm{UV}}$, eq. \ref{eq:beta}. This is easily accomplished in our simulations as we are able to perform monochromatic radiative transfers with relatively high spectral resolutions. However, direct fits are so far impractical on observations since spectroscopic measurements of UV flux at high redshifts are currently unavailable. Thus all UV continuum slopes at high redshifts from observations are computed photometrically from color conversion formulae using two or three bands. Though these formulae carry various justifications to their ability to obtain unbiased estimates of $\beta_{\mathrm{UV}}$, here we take the opportunity to explicitly test one set of examples of color conversion formulae utilized to produce some of the observational results in figure \ref{fig:beta} at $z\sim 5$ - $z\sim 8$ \citep{rei:bio10,rei:biol14}. The results are presented in figure \ref{fig:cc} where red-solid lines are computed from color conversion formulae, blue dashed lines are results of direct least-square fits to the same data from instantly sublimated dust, and bands illustrate the standard error of mean for each magnitude bin. 

The good news is that for all redshifts tested the color conversion formulae appear to be able to estimate UV continuum slopes consistently for all UV magnitude bins. However it is clear that there is a redshift-dependent overall bias towards redder $\beta_{\mathrm{UV}}$ that is more significant at lower redshifts, such that at $z\sim 5$ UV continuum slopes estimated from the color conversion formula are $\sim 0.2$ redder. This bias is greater than the error bars of observations at this redshift, and more or less comparable to the rms band. Thus this bias is not very significant given our simulation volume and galaxy count, but its effect on the $\chi^2$ value is unambiguous. The results shown are only for one version of color conversion formulae and therefore may not be applicable to other versions used in the literature. But they do highlight the possibility of systematic biases in the estimates produced by color conversion formulae. In our particular case the biases are essentially flat across UV magnitudes, and simple corrections can be made. The results from our test illustrate that estimates of $\beta_{\mathrm{UV}}$ produced by color conversion formulae can possess biases greater than the errors, and therefore the accuracy of current observations may be lower than expected. These systematic biases can be controlled by explicitly testing the color conversion formulae with simulations and applying them to data analyses with corrections obtained from the tests.

\end{document}